# Acoustic mode hybridization

# in a single dimer of gold nanoparticles

*Adrien Girard, Hélène Gehan, Alain Mermet, Christophe Bonnet, Jean Lermé, Alice Berthelot,*

*Emmanuel Cottancin, Aurélien Crut, and Jérémie Margueritat\*.*

Institut Lumière Matière, Université de Lyon, Université Claude Bernard Lyon 1, UMR CNRS

5306, F-69622 Villeurbanne, France

ABSTRACT:

**The acoustic vibrations of single monomers and dimers of gold nanoparticles were investigated by measuring for the first time their ultra-low frequency micro-Raman scattering. This experiment provides access not only to the frequency of the detected vibrational modes, but also to their damping rate, which is obscured by inhomogeneous effects in measurements on ensembles of nano-objects. This allows a detailed analysis of the mechanical coupling occurring between two close nanoparticles (mediated by the polymer surrounding them) in the dimer case. Such coupling induces the hybridization of the vibrational modes of each nanoparticle, leading to the appearance in the Raman spectra of two ultra-low frequency modes corresponding to the out-of-phase longitudinal and transverse (with respect to the dimer axis) quasi-translations of the nanoparticles.**

**Additionally, it is also shown to shift the frequency of the quadrupolar modes of the nanoparticles. Experimental results are interpreted using finite-element simulations, which enable the unambiguous identification of the detected modes and, despite the simplifications made, lead to a reasonable reproduction of their measured frequencies and quality factors. The demonstrated feasibility of low frequency Raman scattering experiments on single nano-objects opens up new possibilities to improve the understanding of nanoscale vibrations, this technique being complementary with single nano-object time-resolved spectroscopy as it gives access to different vibrational modes.**

The acoustic vibrational modes of nanoparticles (NPs) reflect their intrinsic properties (size[1,2], shape[3], elasticity[4] and crystallinity[5–7]) and those of their surrounding medium[8–18]. Acoustic measurements thus represent a powerful tool for characterizing both nano-objects and macroscopic materials, nano-objects playing in the latter case the role of nano-opto-acoustic transducers acting as sources and detectors[19]. They can be practically operated using optical approaches in time (e.g., time-resolved spectroscopy) and spectral (e.g., Raman/Brillouin scattering spectroscopy) domains. These techniques enable the detection of a few specific vibrational modes, with however different selection rules making them complementary. They have been intensively used in the last two decades to address the vibrational properties of nano-object assemblies, and in particular the dependence of the acoustic mode frequencies on the properties of the nano-objects[3,20–25], which mostly reflect their intrinsic morphological and elastic properties when they stand in softer solid or low-viscosity liquid environments[26]. These studies have in particular demonstrated the surprising validity of continuum mechanics approaches at the nanoscale, even for ultra-small $\approx 1$ nm diameter NPs[27,28] . The design of time-resolved experiments on single nano-objects[29–31] has constituted an important breakthrough enabling a more detailed investigation of vibrational damping processes, whose study is challenging in ensemble measurements due to the inhomogeneous broadening effects induced by the unavoidable morphological dispersion of the NPs composing the investigated assembly. Measurements on suspended nano-objects in air or liquid have in particular permitted to estimate the quality factors associated with the two mechanisms at the origin of vibrational damping, i.e. the emission of acoustic waves in the environment and the intrinsic processes occurring within the nano-objects[12].

A new direction of research is currently emerging in the field, namely the investigation of the vibrational coupling between close nano-objects occurring when the vibrations generated in a nano-object can propagate to close ones. The simplest system where it may occur is the NP dimer, and its signature was indeed observed in the context of time-resolved experiments on single dimers made of touching NPs, through the appearance of a low-frequency contact-dependent vibrational mode[32].

Initial experimental attempts made to observe acoustic interactions in nanodimers formed by two close but not contacting nano-objects (pairs of nanoprisms and nanocubes lithographed on a substrate) showed no evidence of coupling effects[33,34]. Such evidence could however be recently obtained in two very different systems: ensembles of chemically synthesized monodisperse NPs immersed in a polymer layer[16] (which drastically enhances their acoustic interactions as compared to the previous cases where they could only occur *via* the substrate on which the NPs stand), and single clusters of ~10 close gold nanodisks produced by electron beam lithography[35].

In this paper, we report the first ultra-low frequency (ULF) Raman measurement on single gold nanosystems (isolated NP and nanodimer). This enables us to reproduce our previous ensemble experiments[16] at the single nanodimer level, which provides additional insight on vibrational coupling mechanisms. In particular, we show that the suppression of inhomogeneous broadening not only enables a quantitative analysis of vibrational mode widths, but also the detection of a novel hybridized vibrational mode which had not been considered in our previous analysis, and is expected to be undetectable in the context of time-resolved experiments. Detailed finite-element modeling (FEM) simulations are presented, allowing to ascribe this mode to the hybridization of dipolar vibrational modes with displacements transverse to the dimer axis.

The measurement and detailed analysis of the acoustic properties of a single isolated gold nanosystem by inelastic light scattering were made possible by successively characterizing it with three experimental techniques: Transmission Electron Microscopy (TEM, aiming at the localization and morphological characterization of isolated NPs and dimers), Spatial Modulation Spectroscopy (SMS, providing access to their absolute extinction spectrum) and Ultra Low Frequency Raman Spectroscopy (ULFRS, yielding their inelastic scattering spectrum). Another key to the success of these experiments was the choice of a rather large NP diameter (about 100 nm), providing signals sufficiently large for achieving single-NP sensitivity despite the low efficiency of the Raman scattering process, in a ULF range (<15 GHz) still accessible with our Raman spectroscopy apparatus.

Figure 1 shows the different steps followed in our studies. Gold NPs were chemically synthesized using a seeded growth technique already reported in the literature[16]. The resulting NPs were stabilized and transferred in ethanol using polyvinylpyrrolidone (PVP). A 5µL drop of this highly diluted solution was then deposited onto the 40 nm silica membrane of a TEM substrate and left dry several minutes (Figure 1a). In a second step, rapid low magnification TEM images were taken to locate nanosystems of interest without damaging them[36]. White light SMS (whose detailed description can be found elsewhere[37]) images were then acquired and compared to the TEM ones to retrieve the nano-objects of interest, and subsequently acquire their absolute extinction cross-section spectra (Figure 1c). Then, their ULFRS was performed. The microscopy grid was placed on top of an inverted microscope, with the 647 nm line from a Krypton laser focused with a x100 microscope objective used as the excitation beam (Figure 1d). A light power of about 0.5 mW at the level of the sample was used (corresponding to the illumination of the investigated nanosystem with an intensity of the order of 1 mW/µm$^2$), allowing to perform long acquisitions (up to several days) without damaging the NPs. Rayleigh scattering maps were acquired and compared with TEM and SMS images. The laser spot was positioned onto the object of interest and the inelastic light scattering signal was collected by a Brillouin spectrometer (Tandem Fabry-Perot from JRS Scientific Instruments) with a spectral resolution of about 50 MHz. For the experiments reported here, the free spectral range of the spectrometer was fixed to 30 GHz (i.e., frequency shifts from -15 GHz to 15 GHz were measured), allowing resolution of vibrational modes down to 3 GHz frequencies. The typical acquisition times for these experiments ranged between 2 and 6 hours, depending on the optical response of the investigated system. Drift issues during the acquisitions were minimized by the use of a capacitive piezo-electric translation stage with high stability over time, the permanent realignment of the spectrometer (using a small portion of the incident laser line) and the placement of the whole experimental system (microscope and Brillouin spectrometer) on a piezo active isolator. High resolution TEM microscopy of the investigated nanosystems was finally performed in order to precisely characterize their geometry and in particular their size, which constitutes a crucial parameter for the FEM simulations since the vibrational frequencies of a NP scale as the inverse of its size.

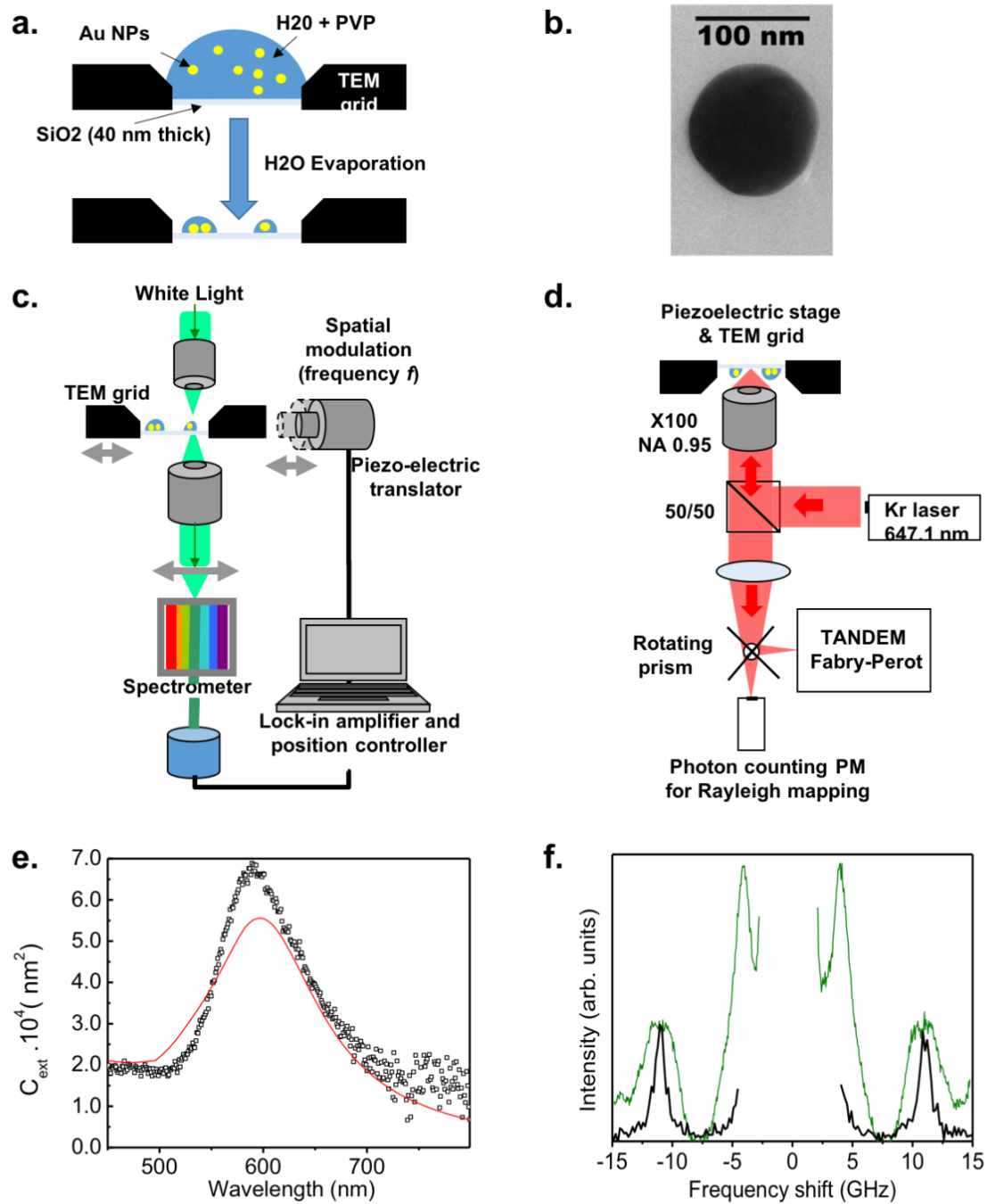

**Figure 1: a.** TEM grid preparation to obtain monomers and oligomers of polymer-embedded gold NPs. **b.** High resolution TEM image of an isolated NP, obtained after all the optical measurements to avoid any structural modification due to the electron beam. **c.** Experimental setup used for the SMS measurements. **d.** Experimental setup used for ULFRS and Rayleigh mapping. **e.** Extinction cross-section of the monomer measured by SMS (black dot curve), compared to the theoretical extinction spectrum of a 96 nm diameter gold nanosphere immersed in a PVP matrix (n = 1.5), obtained using Mie theory (red line). **f.** ULFRS spectrum of the NP shown in panel **b** (black line). The spectrum obtained on an assembly of NPs embedded in PVP matrix (as described in ref. 16) is shown by the green curve for comparison. The central line corresponding to Rayleigh scattering has been omitted for clarity.

Figure 1b represents the TEM image of the single isolated NP with a 96 nm average diameter (determined from the electron microscopy image with an accuracy of about 2 nm) whose acoustic response is presented in the following. The presence of a PVP layer around the NP cannot be directly confirmed by this image. However, it is suggested by the comparison of the absolute NP extinction cross-section spectrum measured by SMS (black dots in Figure 1e) with the one calculated using Mie theory considering a 96 nm diameter gold nanosphere immersed in a homogeneous medium of 1.5 refractive index, i.e. that of PVP (red line in Figure 1e). Indeed, both spectra show a peak associated with the dipolar localized surface plasmon resonance of the NP, with central positions close for the measured and computed spectra (about 600 nm; in contrast, Mie calculations with an environment refractive index of 1.2 – an effective value modeling the inhomogeneous environment of a NP standing in air on a glass substrate – lead to a very different 550 nm central position). The measured optical response is thus consistent with the presence of a thick PVP layer around the NP. Note that ULFRS experiments benefit here from the red-shift of the surface plasmon resonance wavelength resulting from retardation effects (induced by the large NP size) and high refractive index local environment (due to the PVP matrix), which permits to work in quasi-resonant conditions with the 647 nm excitation wavelength used and thus increases the internal field driving the inelastic Raman scattering[20].

Figure 1f compares the ULF Raman spectrum obtained on the single NP shown in Figure 1b (black curve) to that obtained on an assembly of NPs synthesized using the same procedure (green curve). Both spectra display a peak at $f_{NP}$=11.0 GHz, ascribed to the quadrupolar vibrational mode (corresponding to a $\ell = 2$ angular momentum number) of the NPs, which usually dominates their Raman scattering spectra. This interpretation is confirmed by the prediction by Lamb theory[38] of a 10.5 GHz frequency for this mode in the case of a 96 nm diameter homogeneous gold nanosphere in vacuum (including the PVP environment in the calculations[39] leads to a very close 10.6 GHz frequency value). However, this peak is three times narrower in the single NP case as compared to the assembly one. This effect is related to the suppression of inhomogeneous broadening (i.e., the fact that the NPs of an assembly vibrate at different frequencies due to the dispersion of their geometrical characteristics) in single-particle experiments, which is one of their major advantages as it enables to quantitatively investigate the physical mechanisms governing vibrational damping[12,40]. This narrowing will be further discussed in the following. Another difference between the two spectra shown in Figure 1f is the presence of a ULF peak at 4.0 GHz in the assembly case, absent in the single-particle one. In the context of a previous study on NP ensembles[16], this peak was ascribed to the fraction of the NPs sufficiently close from each other to mechanically interact as dimers, or possibly more complex oligomers. This interpretation was supported by a theoretical modeling of the vibrational modes of dimers of NPs mechanically coupled *via* a polymer matrix, showing that hybridization of their quasi-translation $\ell = 1$ vibrational modes occurs, yielding in particular a coupled mode that is Raman-active, in contrast with the unhybridized $\ell = 1$ modes[16]. The measurements shown in Figure 1f provide additional support to this interpretation by showing the absence of this ULF mode for a single isolated NP.

The interpretation of our previous ensemble measurements[16] can be tested even more directly by considering the simplest nanosystem leading to inter-particle mechanical interactions, i.e. a dimer of two close NPs surrounded by a polymer layer. The experimental results obtained with a single isolated dimer, as well as their analysis by FEM simulations, are presented in Figure 2.

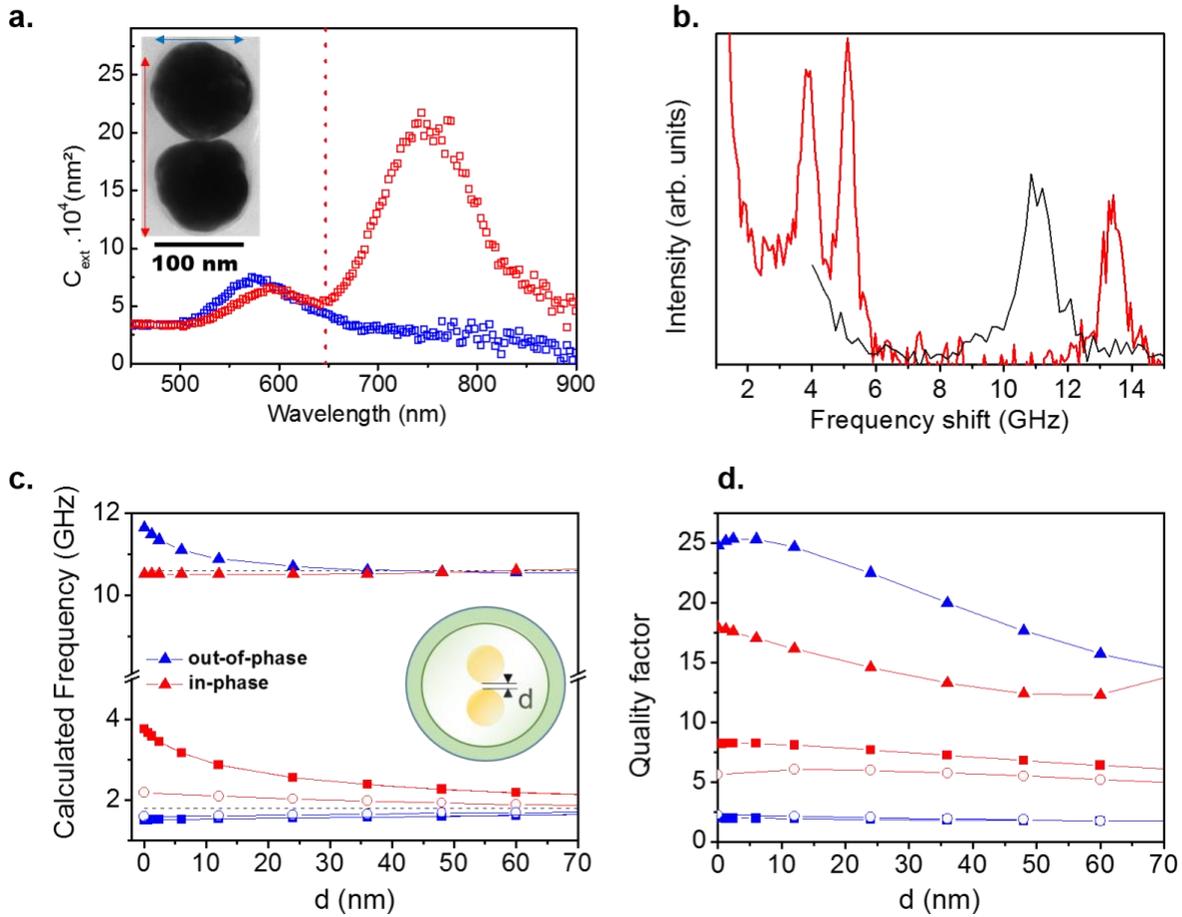

**Figure 2: a.** Absolute extinction cross-section of a single dimer for incident light polarized along its short (blue dots) and long axis (red dots). The excitation wavelength used in ULFRS experiments is shown by the vertical dotted line. The inset represents the TEM image of the dimer studied. **b.** Raman scattering spectrum of the dimer excited with a polarization parallel to its long

axis (red curve). The spectrum obtained with the monomer is shown for comparison (black curve, same as in Figure 1f). **c.** FEM-computed frequencies of the acoustic modes of a dimer of two 96 nm diameter nanospheres in an infinite PVP matrix, as a function of their separating distance $d$. The two modes generated by the hybridization of the ($\ell = 2; m = 0$) modes of individual NPs are shown by triangles. Those produced by the hybridization of longitudinal (i.e., with displacement mostly along dimer axis) and transverse (displacement mostly orthogonal to dimer axis) $\ell = 1$ modes are shown by squares and hollow circles, respectively. In each case, the hybridization generates two types of modes corresponding to in-phase and out-of-phase vibrations of each NP, indicated by blue and red symbols, respectively. The horizontal dashed lines show the frequencies of the $\ell = 1$ and $\ell = 2$ modes of isolated NPs. The inset represents the geometry used in the numerical calculations (the green shell represents the perfectly matched layer). **d.** Calculated quality factors (same symbols as in panel **c**).

The inset of Figure 2a shows the dimer studied, constituted by two NPs with an average diameter of 96 nm. The absolute extinction spectra of this dimer measured by SMS for incident light polarization parallel or orthogonal to the dimer axis are also shown in Figure 2a. For orthogonal polarization (blue dots), a peak at 570 nm is observed. It corresponds to the transverse dipolar plasmon mode of the dimer, whose position is slightly blue-shifted with respect to that of the dipolar plasmon modes of an isolated sphere (figure 1e). Such shift can be understood as a result of the hybridization of dipolar plasmonic modes[41,42]. For light polarization along the dimer axis, two peaks with 760 nm and 595 nm central positions are observed. The first one, at 760 nm, corresponds to the main longitudinal hybridized dipolar plasmon mode of the dimer. The second one at 597 nm corresponds to the hybridized quadrupolar plasmon mode of the dimer, which has acquired a noticeable dipolar character through the hybridization mechanism (mixing of isolated

NP modes of different angular momenta occurring in the strong coupling regime), explaining its large extinction cross-section over a wide spectral range.

ULFRS measurements were performed with 647 nm excitation (dotted line in Figure 2a), i.e. between the two plasmonic modes of the dimer described above. The ULFRS spectrum measured for this dimer (red curve in Figure 2b) is considerably richer than that acquired on a single NP (black curve in Figure 2b) and presents three different peaks with frequencies $f_{1dimer}$=13.4 GHz, $f_{2dimer}$=5.2 GHz and $f_{3dimer}$=3.9 GHz. The measured $f_{1dimer}$ frequency is about 25% higher than those computed for the quadrupolar vibrational modes of the two 96 nm gold NPs forming the dimer (10.5 GHz and 10.6 GHz for NPs in vacuum and PVP, respectively, as discussed above). The most likely explanation for this frequency shift is the hybridization of the quadrupolar modes of the NPs forming the dimer. To confirm this hypothesis, we performed FEM simulations of the acoustic modes of a dimer of mechanically coupled NPs, exploring the effect of their mechanical coupling by varying their separating distance. An accurate quantitative modeling of the experiments is difficult for several reasons, including the imperfectly spherical shape of the NPs used, their complex, inhomogeneous environment (due to their deposition on a substrate and the finite PVP layer surrounding them), uncertainties on the inter-NP and NP-substrate distances, and the five-fold degeneracy of the quadrupolar mode for elastically isotropic nanospheres, making their hybridization potentially complex. Therefore, we limited ourselves to a qualitative description of hybridization effects, considering the ideal case of gold nanospheres in an infinite PVP matrix, additionally assuming axial symmetry of the displacement fields about the dimer axis (which corresponds to considering only the hybridization of the isolated NP modes with an $m = 0$ azimuthal number).

These calculations, presented in Figure 2c, show for the angular momentum $\ell = 2$ a frequency splitting increasing with decreasing interparticle distance. Among the two hybridized acoustic modes, the highest-frequency one, corresponding to an in-phase combination of the $\ell = 2$ NP modes, induces a larger modulation of the interparticle distance than the out-of-phase combination at lower frequency, and is thus expected to show a much stronger Raman activity, explaining the detection of a single quadrupolar peak in the experiments. For very close NPs (in the regime where their separating distance is much smaller than their diameter; note that vibrational parameters have well-defined finite limits in this regime, making precise knowledge of gap distance less crucial in these acoustic studies than in the context of optical ones), its frequency exceeds by $\approx 10\%$ that of the quadrupolar mode of an isolated NP. Such hybridization of quadrupolar modes had never been detected before, neither with ULFRS on assemblies[16] (the $\approx 15\%$ frequency shift that it induces being of the order of the inhomogeneous broadening caused by the dispersion of NP diameters and inter-NPs distances and moreover affecting only a fraction of the NPs) nor using time-resolved spectroscopy (as the quadrupolar modes of nanospheres are not excited with this technique[12]).

The observation in the ULF domain of two peaks at $f_{2dimer}$=5.2 GHz and $f_{3dimer}$=3.9 GHz (Figure 2b) contrasts with that of a single, broader one at 4.0 GHz for NP assemblies (Figure 1f). This latter peak was interpreted as the result of the hybridization of the dipolar translation-like ($\ell = 1$) vibrational modes of NPs in dimers[16], in connection with axisymmetric (i.e., considering only m=0 vibrational modes) FEM simulations leading to a good reproduction of its frequency[16]. The reduction of mode width associated with single-particle experiments demonstrates that hybridization phenomena actually generate two distinct vibrational modes, probably made undistinguishable in ensemble measurements because of their overlap induced by inhomogeneous broadening. We ascribe the $f_{2dimer}$=5.2 GHz and $f_{3dimer}$=3.9 GHz peaks to the Raman-active combinations of $\ell = 1$ modes involving displacements parallel and orthogonal to the dimer axis, respectively. This conclusion was supported by the combination of axisymmetric 2D FEM simulations similar to those described in ref. 16 (yielding uncoupled and coupled $\ell = 1$ modes with displacement parallel to the dimer axis) with 3D ones also providing access to modes with displacements orthogonal to the dimer axis. This analysis leads to a total of four different types of hybridized modes (Figure 3). The two modes with the lower frequencies approximately correspond to the in-phase translation of each NP, and thus to a global translation of the whole dimer. Therefore, they are not expected to strongly modulate its optical response, and their Raman activity is thus expected to be weak. In contrast, the two highest frequency ones, corresponding to out-of-phase combinations, are expected to inelastically scatter light much more efficiently as they induce a strong modulation of the interparticle gap (Figure 3) and thus of the dimer scattering properties[43]. Their 3.8 and 2.2 GHz frequencies computed for 96 nm diameter NPs in contact are of the same order of magnitude as $f_{2dimer}$ and $f_{3dimer}$. The differences are probably due to the necessary oversimplification of the ideal system (geometry and mechanical properties) performed for carrying out the FEM simulations. Such hybridization of acoustic dipolar modes presents a strong analogy with the hybridization of dipolar plasmonic modes in homodimers of NPs, which generates four distinct coupled modes, among which two are significantly coupled with light: a red-shifted one corresponding to plasmon oscillation along the dimer axis, and a blue-shifted (by a smaller amplitude) one associated with an oscillation direction orthogonal to the dimer axis[42,44–47]. These hybridized plasmonic modes can be selectively excited by aligning light polarization parallel or orthogonally to the dimer axis, which is not the case of vibrational modes in the context of Raman spectroscopy, based on their thermal excitation.

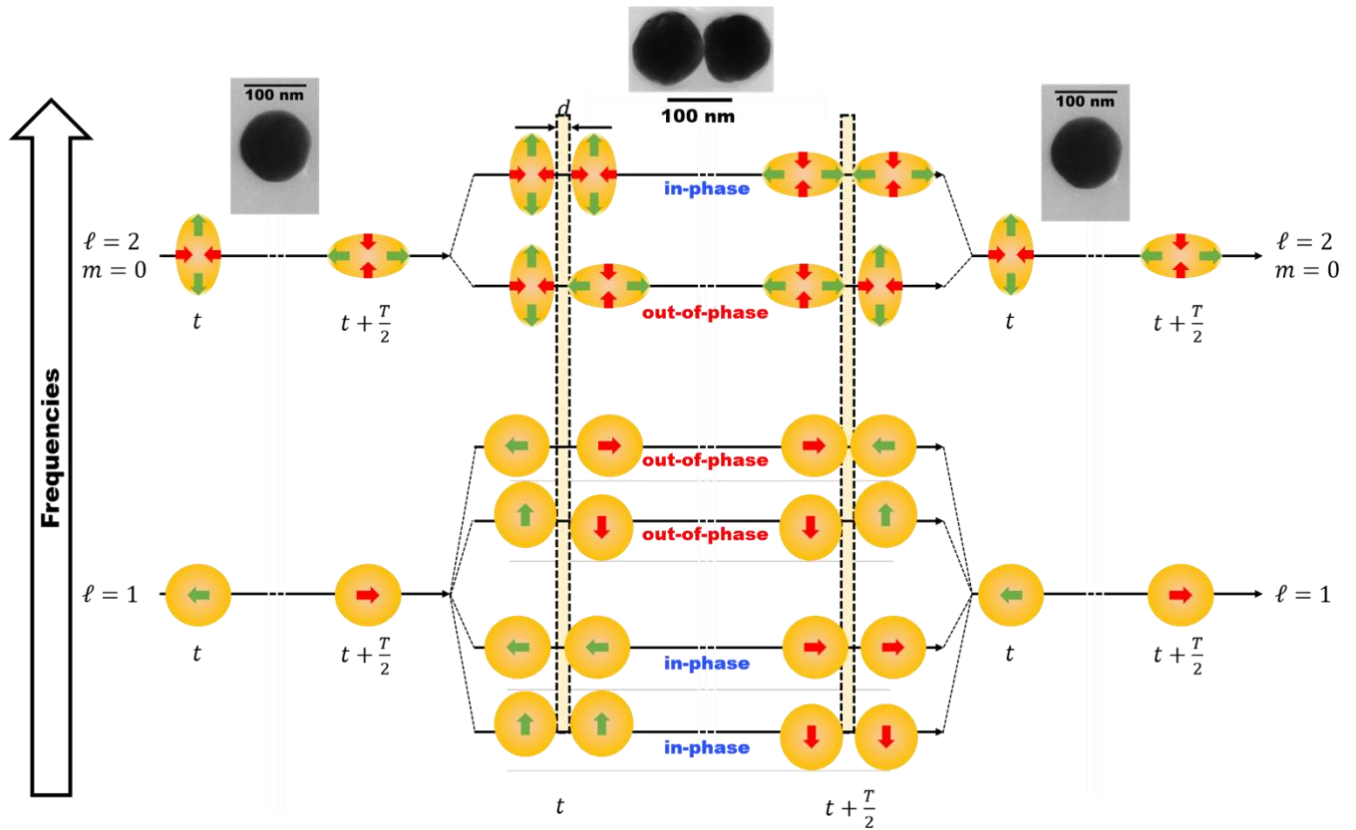

**Figure 3:** Schematic description of the hybridization of the translation-like ($\ell = 1$) and quadrupolar ($\ell = 2$) vibrational modes in a dimer of close NPs. All vibrational modes are represented at two instants t and t+T/2 differing by half of their associated period T, with displacement amplitudes intentionally made much larger than those induced by thermal motion at ambient temperature for clarity. The $\ell = 1$ mode is represented by a simple translation, ignoring the small NP deformation that it induces. Only the quadrupolar mode with m=0 is considered on the figure. It is represented by a modulation of the NP shape, i.e. an increase of its length in one direction concomitant with a decrease of its diameter in the transverse plane. The dashed rectangles represent the interparticle distance in the absence of vibration and schematically underline the potential of each hybridized mode to modulate the interparticle distance.

As a result of the suppression of the inhomogeneous broadening contribution in our single-particle experiments, acoustic modes appear as much narrower (as compared to the ensemble case) resonances in Raman spectra (Figure 1f), whose widths reflect their homogeneous damping rate (this is the equivalent in the spectral domain of the increased lifetime of signal oscillations that is usually observed with time-resolved experiments performed on single nano-objects[26,48]). The quality factors of the detected vibrational modes of isolated NPs and nanodimers were deduced from their Raman scattering spectra (Figure 2b) as Q=f/Γ, with f the central frequency of the resonances and Γ their full width at half maximum. This analysis yields $Q_{NP}$=9±1 for the quadrupolar mode of the isolated NP. In the dimer case, $Q_{1dimer}^{exp}$ = 18±2, $Q_{2dimer}^{exp}$= 9±1 and $Q_{3dimer}^{exp}$= 5±1 values were obtained for the detected modes respectively associated with the hybridized quadrupolar modes and longitudinal and transverse quasi-translation ones. These measured quality factors are of the same order as the 5 to 50 ones previously measured by single-particle time-resolved spectroscopy for the breathing and extensional modes of spherical and elongated isolated nano-objects deposited on a substrate[26].

They were compared to those predicted by our FEM model for NPs and dimers in an infinite PVP matrix (Figure 2d). It should be noted that in the context of FEM calculations, the spectral widths computed for the vibrational modes are much more sensitive to simulation parameters (mesh size, perfectly matched layer position and thickness) than their central frequencies. However, the availability of analytical results in the matrix-embedded sphere case[39,49] allowed us to determine the simulation conditions providing reliable results. Our simulations show that hybridization effects have a strong impact on vibrational quality factors, even for interparticle separating distances of the order of their diameter (Figure 2d). In particular, a significant improvement of the quality factors of all the Raman-active dimer modes is predicted in the quasi-contact regime ($d \ll D_{Au}$). In this range, our model predicts $Q_{3dimer} < Q_{2dimer} < Q_{NP} < Q_{1dimer}$, which corresponds well to the experimental observations. Moreover, the computed quality factor values ($Q_{NP}^{theo}$=14, $Q_{1dimer}^{theo}$=25, $Q_{2dimer}^{theo}$=8 and $Q_{3dimer}^{theo}$=6) are in reasonable agreement with the measured ones despite the strong simplifications made in the modeling, and notably the use of an infinite PVP matrix around the NPs.

In conclusion, ULFRS measurements were performed for the first time on individual gold nanosystems. Interpretation of the obtained inelastic scattering spectra by FEM calculations (performed with a simplified geometry) revealed fine details about the hybridization of acoustic modes in a single isolated dimer, inaccessible in previous ensemble experiments. First, the hybridization of the quasi-translation ($\ell$ =1) modes of the NPs forming the dimer (Raman-inactive for isolated NPs) was demonstrated to generate two detectable modes with distinct frequencies, respectively associated with longitudinal and transverse out-of-phase motions of both NPs. Acoustic mode hybridization in a NP dimer was also shown to increase the frequency of the detected peak associated with in-phase quadrupolar ($\ell$ =2) modes, which dominate Raman scattering by isolated NPs. Finally, hybridization was experimentally demonstrated to significantly modify the quality factors of acoustic modes, in qualitative agreement with FEM calculations.

Vibrational coupling effects therefore appear as a promising way to tune the properties of NP acoustic modes. A strong parallel can be drawn here with the topic of interacting NP plasmon modes, which have been intensely investigated from experimental, theoretical and technological perspectives in the last two decades, and have emerged as a powerful means to control electromagnetic radiation at the nanoscale[47]. However, acoustic interactions constitute a much newer field of research and are currently much less understood than plasmonic ones. Their finer characterization will require systematic investigations on NP dimers (e.g., experimentally checking the effects of the inter-NP distance and the properties of their coupling medium) and consideration of NP oligomers formed by more than two NPs (as already started in the context of a recent study[35]), similarly to what has been done on plasmonic modes. The detailed and quantitative information provided by single-particle ULFRS investigations should also enable a better understanding of Raman scattering by single nanosystems, i.e. of the coupling between their acoustic and optical responses. In particular, our investigations on NP dimers (i.e., this study and the one described in ref. 16) have demonstrated a spectacular enhancement, and shows that the vibrational hybridization induced by mechanical coupling makes Raman active vibrational modes that are not observed in the case of isolated NPs. In this work, the efficiency with which a dimer acoustic mode inelastically scatters light was qualitatively related to the modulation of the inter-NP gap that it induces. Development of theoretical models of inelastic light scattering by single and interacting plasmonic NPs will be required for a more quantitative analysis of the amplitude of experimental ULFRS spectra.

AUTHOR INFORMATION

**Corresponding Author**

*jeremie.margueritat@univ-lyon1.fr, Institut Lumière Matière-(ILM)-UMR5306, Bâtiment Alfred Kastler, 10 rue Ada Byron, 69622 Villeurbanne cedex—France.

ORCID :

Jérémie Margueritat : 0000-0003-2075-1875

**Author Contributions**

A.M and J.M got the idea of the measurements and J.M designed the ULFR experiment. H.G synthesized the nanoparticles. A.G performed the ULFR measurements, the SMS measurements, and the TEM measurements with J.M, C.B, E.C. The analysis of the experimental data was performed by A.G, A.B, J.L, A.C, and J.M, and A.B, A.C, and J.M wrote the manuscript, to which all the authors contributed via discussions and corrections. The authors declare no competing interest.

**ACKNOWLEDGMENT**

This work was supported by the ANR NanoVip project, Grant ANR.13.JS10.0002 of the French Agence National de la Recherche and the Fédération André Marie Ampère 2013 (FRAMA). The authors declare no competing financial interest. We thank the Centre Technologique des Microstructures at Villeurbanne (CTμ) for access to the microscope platform TEM characterizations.